\documentclass[a4paper,12pt]{article}

\usepackage[austrian,american]{babel}
\usepackage[intlimits,leqno]{amsmath}
\usepackage{amssymb} 
\usepackage{amsthm}
\usepackage{graphicx,color}
\usepackage{makeidx}
\usepackage{epsfig}
\usepackage{latexsym}
\usepackage{pstricks,pst-node,amsfonts}

\usepackage{cancel}

\theoremstyle{plain}

\newtheorem{rema}{Remark}

\newtheorem{exam}{Example}

\newcommand\oA{{\mathcal{A}}}
\newcommand\oB{{\mathcal{B}}}
 
\newcommand\oP{{\mathcal{P}}}
\newcommand\oQ{{\mathcal{Q}}}
\newcommand\oT{{\mathcal{T}}}
\renewcommand\H{{\mathcal{H}}}

\newcommand\N{{\mathbb{N}}}
\newcommand\R{{\mathbb{R}}}

\newcommand{\A}{\mathcal{A}}

\newcommand{\X}{\mathcal{X}}
\newcommand{\Y}{\mathcal{Y}}
\renewcommand{\L}{\mathcal{L}}
\newcommand{\F}{\mathcal{F}}

\newcommand{\D}{\mathcal{D}} 
\renewcommand{\i}{{\rm i}}

\newcommand{\x}{\mathbf{x}}

\renewcommand{\u}{\mathbf{u}}
\renewcommand{\v}{\mathbf{v}}

\renewcommand{\d}{{\rm d}}

\newcommand\Id{ \mbox{\,Id\,} }

\begin{document}

\providecommand{\keywords}[1]
{
  \small	
  \textbf{\textit{Keywords:}} #1
}

\title{Extension of the principle of least action with focus on dissipative equations}

\author{ Richard Kowar\footnote{Department of Mathematics, University of Innsbruck, 
Technikerstrasse 21a, A-6020, Innsbruck, Austria}
\footnote{richard.kowar@uibk.ac.at}
}

\maketitle

\begin{abstract}
In this paper, we extend the \emph{principle of least action} and show that a \emph{Lagrange density} 
always exists for the usual linear pde or linear fractional problems $\oA\,u=f$ in physics, if the usual 
causality conditions $u|_{t<0}=0$ and $f|_{t<0}=0$ are assumed. (The approach is actually applicable to 
uniquely solvable linear operator equations for which an adjoint exist.) The set of Lagrange densities 
together with the zero vector form a non-trivial vector space and for each different set of variables, e.g. 
$\{u_t,f\}$, $\{u_{xt},f\}$ or $\{u_t,u_x,u_y,u_z,f\}$, there exists a Lagrange density 
that implies a Lagrange equation, which is equivalent to the considered problem. 
The usual Lagrange density is such that it implies the 'original equation'. But there are pde's for which 
the standard theory does not imply a Lagrange density. Such equations are e.g the 
\emph{advection-diffusion equation}, 
the \emph{shear wave equation}, 
the \emph{Telegraph equation} and 
the \emph{Nachman, Smith and Waag wave equation}. 
We show that for each of these equations a (covariant) Lagrange density exists that leads to an equivalent 
\emph{higher order pde} (if it is formulated with the above causality conditions). For each of these equations, there 
exists a Lagrange density that implies a Lagrange equation that equals  the original equation, but this 
Lagrange density contains at least one \emph{linear integral operator} (actually an inverse of a partial 
differential operator). 
A new point of view is that each of these equivalent Lagrange densities for a given set of variables  
implies a (usually different) \emph{generalized Hamiltonian density}, where the respective 'Hamiltonian' 
is conserved if $\oA$ and $f$ are appropriate. 
The standard Lagrange density implies an Hamiltonian that (frequently) models the energy. Morever, each 
conserved Hamiltonian implies countable many higher order Hamiltonians that are conserved (if the solution 
of the considered problem is sufficiently smooth.) For a better understanding, several concrete examples 
have been included in various places.  
\end{abstract}

\keywords{Principle of least action, Lagrange equations, Lagrange density, Hamiltonian density, 
dissipative equations, Nachman-Smith-and-Waag equation}


\section{Introduction}
\label{sec-intro}

The \emph{principle of least action}\footnote{It would be more appropriate to call it the principle of optimal action.} 
is a powerful and successful 'tool' for modeling, characterizing and analysing problems in science. 
Especially, if the Lagrange density is \emph{covariant} (cf.~\cite{SuHr13,SuFr20a,SuFr20b}). 
The significance of the principle of least action is well known for many (non-dissipative) processes  
(cf.~\cite{FeLeSa63,LanLif91me,FeWa80,Jac99,Evans98,Sc89,GrRe93,LanLif91el,
LanLif91hy}). It states that an equation, say $\oA\,u=f$, can be obtained by optimizing 
the \emph{action} 
$$
   u\mapsto S_f[u] := \int_\Omega  \L[u,f](\x,t) \,\d \x \,\d t \qquad\quad 
$$ 
over the vector space $\D  := C_0^\infty(\Omega)$ with $\Omega :=\R^3\times I$, where $I\subseteq \R$ 
is an approrpiate time interval.  
The respective \emph{optimality condition} is given by $S_f'[u] h = 0$ for $h\in\D$, 
where $S_f'[u]h := \lim_{s\to 0}\frac{1}{s}\left( S_f[u+sh] - S_f[u] \right)$ denotes the 
\emph{Gateaux derivative} of the action $S_f$ at $u$ in 'direction' $h$. If the principle 
is applicable, then the optimality condition leads to (cf.~\cite{FeLeSa63,BlBr92})
$$
       \int_\Omega \left( \oA u - f \right)\,h \,\d \x \,\d t = 0 \quad\mbox{ for } \quad h\in\D
                   \qquad\mbox{ and thus }\qquad       
       \oA\,u=f \quad \mbox{on $\Omega$}\,.
$$ 
For many applications, a \emph{Lagrange density} $\L$ is equal to the \emph{kinetic energy density} minus 
the \emph{potential energy density}. In field theory, it is common to assume that $\L$ is a function of
 $u_t$, $u_x$, $u_y$, $u_z$, $u$ which leads to the formal Lagrange equation 
$\partial_\mu\,(\partial_{\partial^\mu u} \L) + \partial_u \L = 0$, 
where $\partial^0 := -\partial_0 := \frac{1}{c}\,\partial_t$, 
$\partial^1 := \partial_1 :=\partial_x$, ..., $\partial^3 := \partial_3 := \partial_z$.

\subsection{Motivation}

This paper was motivated from the fact that Lagrange densities of dissipative equations like 
the \emph{advection-diffusion equation}, the \emph{Telegraph equation} and the \emph{Nachman, Smith and Waag 
wave equation} do not appear in the literature. As a consequence, we tried to derive Lagrange densities for these 
equations. Actually, the derived Lagrange densities implied Lagrange equations that where pde's of an higher order,   
but each of them (with the right source term $f$ and the causality conditions) where equivalent to the original 
problem. After that we did the same for other equations like Airy's equation, the Beam equation 
and the shear wave equation. (The latter equation is 'similar' to the diffusion equation.) This first approach is also 
applicable to special time fractional equations (of rational order), as will be demonstrated later.  
For all this, it was crucial to work with the \emph{action functional} (similar as in~\cite{FeLeSa63}) instead 
of the \emph{formal Lagrange equation} that assumes a special dependency of the Lagrange density. 
Then we realized that there are also Lagrange densities that depend on the usual variables $u_t$, $u_x$ ,..., $u$, 
but they contain at least one linear integral operator (actually an inverse of a partial differential operator). 
As a consequence, we reassessed our first approach and realized that all equivalent Lagrange 
densities\footnote{Equivalent means that they imply an equation that is equivalent to the original one.} 
are meaningful. They all have to tell us something. This led us  to an extended approach that is much more 
unitary, but also more abstract. 
The most crucial point is that there are many equivalent Lagrange densities (for a given problem) 
and each of them provides us a with an Hamiltonian density. Which of the respective Hamiltonians are 
conserved depends on the governing equation. So we have 'just' to choose those (equivalent) 
Lagrange densities that are simple, covariant and lead to conserved Hamiltonians. (Actually the other 
ones tell us the same, but it is much more difficult to figure out.) The usual principle of least action focuses  
only on a small part of this larger concept and as a consequence for some problems Lagrange densities 
are missed out or seem not to exist. 
We note that (under reasonable assumptions) there always exist covariant Lagrange densities, but 
this is not true for the Hamiltonian densities. In this sense the Lagrange densities are more important 
than the Hamiltonian densities. Nevertheless, if we have all the Hamiltonian densities, it is possible to construct 
all Lagrange densities. (We do not prove this here, but the reader will find this statement at the end of 
this paper evident.) 

We believe that a new field of research is opened and the expertise from several different areas in physics 
as well as mathematics will shine a brighter light on it. Let us start with it.  \\

The paper is organized as follows. In Section~\ref{sec-basics}, we specify the class of problems that 
we consider and show the existence of Lagrange densities for them. Then we show that for each problem 
the set of Lagrange densities together with the zero vector form a vector space. In this section, we also 
generalize the concept of Hamiltonians and higher order Hamiltonians, and discuss it with the help of several 
examples. Afterwards, in Section~\ref{sec-CaseEx}, we discuss the principle of least action for Airy's equation 
and frequency dependent dissipative wave equations, and discuss the modeling of initial conditions via source 
terms. Special Lagrange densities (that contain only partial differential operators) are introduced and discussed 
in Section~\ref{sec-pdoL} (this was basically our first approach). With our approach, covariant Lagrange densities 
are derived for the Telegraph equation, the Airy equation, the advection-diffusion equation and the Nachman, 
Smith and Waag equation in Section~\ref{sec-covL}.  Further examples like the shear wave equation are 
presented and discussed in Section~\ref{sec-FurEx}.  Finally, in Section~\ref{sec-Concl}, we summarized our 
results and state our conclusion.

\section{The basics}
\label{sec-basics}
\subsection{The considered class of problems}

If not otherwise specified, we assume the following in this paper. 
Let $\oA:\X \to \Y$ be a linear (injective) partial differential operator (pdo) and $\oA^*$ its adjoint. We do not 
assume that the coefficients of the pde are constants. We consider uniquely solvable problems of the type 
\begin{equation}\label{opeq1}
    \oA\,u=f  \quad \mbox{on $\R^4$} 
           \qquad\mbox{with}\qquad 
    u|_{t<0} = 0  
           \quad\mbox{and}\quad 
    f|_{t<0} = 0 \, 
\end{equation} 
or the corresponding initial value problem. The last two conditions are the usual causality conditions. We note that 
the \emph{initial conditions} $(\partial_t^m u)(\x,0+) = u_m(x)$ for $m\in\{0,\,1,\,\dots,\,n-1\}$ can be modeled by 
\begin{equation}\label{deffInit}
       f(\x,t) = \sum_{j=0}^{n-1} u_j(\x)\,\delta^{(n-1+j)}(t) 
                 \quad \mbox{,where}\quad 
       \delta^{(0)} := \delta \quad\mbox{(delta distribution)}\,.
\end{equation}
We tacitly assume that problem~(\ref{opeq1}) is non-dimensionalized. 
This formulation permits us to use the Fourier transform such that for example the uniqueness of the solution 
can be shown.  

Actually, $\oA$ need not be a pdo, as long as it has an adjoint and $\oA^{-1}$ exists. So it may be a 
fractional operator. Even $\oA^{-1}$ need not exist for each part of our approach, but it simplifies and clarifies 
the matter very much.

\subsection{Existence and 'basis Lagrange densities'}

According to our assumptions, there exists an inverse $\oA^{-1}$ (that can be calculated via the Fourier transform) 
such that $u = \oA^{-1}\,f$. Then a Lagrange density of the equation $u = \oA^{-1}\,f$ is given by 
\begin{equation}\label{Ltrivial}
     \L_{triv} = \frac{1}{2}\,u^2 - (\oA^{-1}\,f)\,u  \,.
\end{equation} 
Indeed, $h=\oA^*\,k\in\D:= C_0^\infty(\Omega)$ if and only if $k\in\D$ and thus $0 = S_f'[u](h) $ for all 
$h\in\D$ implies $0 = \int_\Omega \left( u - \oA^{-1}\,f \right)\,h \,\d \x\,\d t
= \int_\Omega \left( u - \oA^{-1}\,f \right)\,(\oA^*\,k) \,\d \x\,\d t 
=\int_\Omega \left( \oA\,u - f \right)\,k \,\d \x\,\d t$ for all $k\in\D$. 
Because the equation $u = \oA^{-1}\,f$ is equivalent to $\oA\,u = f$, this is also a Lagrange density of the original 
problem. 
Moreover, because $u$ is invariant (scalar), $\oA^{-1}\,f$ ($=u$) is also invariant, and thus 
$\L_{triv}$ is also invariant.  
Similarly, if $\oP$ is a linear bounded operator (e.g. $\oP=\partial_\mu$ or  $\oP=\oA$),\footnote{If $\oP=\oA$, 
then $\oA^{-1}$ need not exist and $\L_{\oA}:= \frac{1}{2}\,(\oA\,u)^2 - (\oA^*\,f)\,u$ implies the normal 
equation that is equivalent to the original equation if $\oA$ is injective. 
} then 
\begin{equation}\label{LP}
   \L_{\oP}
       := \frac{1}{2}\,(\oP\,u)^2 - (\oP^*\,\oP\,\A^{-1} \,f)\,u         
   \qquad (\mbox{$\oP^*$ adjoint of $\oP$})
\end{equation}
is also a Lagrange density of problem~(\ref{opeq1}). 
Let us look at~(\ref{LP}) from a different angle. If we are concerned with the wave equation, i.e. 
$\oA := \partial_\mu\,\partial^\mu$, then the classical Lagrange density $\L$ satisfies
$$
    \L
       = -\frac{1}{2}\,\partial_\mu\partial^\mu\,u + f\,u 
       =  \L_{\partial_0^2} - \L_{\partial_1^2} - \L_{\partial_2^2} - \L_{\partial_3^2}\,,
$$
i.e. the 'nasty' terms in the latter sum cancel out and the standard Lagrange density remains.

\begin{rema}
So far, our point of view is that there are many equivalent Lagrange densities that have different formal dependencies,  
say on $u_t$, $u_x$, $\oP\,u$, $u$, and so on. By equivalent, we mean that if we consider the Lagrange densities only as 
function of $u$, then their action implies the considered equation or an equivalent equation. 
The various possible dependencies in the equivalent Lagrange densities determine the form in which the Lagrange 
equation turns up. Some of them are written in covariant form. 
Moreover, it will turn out that each (form of the) Lagrange density determines at least one 
Hamiltonian density. Whether the respective Hamiltonian is conserved depends on the problem. 
\end{rema}

\subsection{Vector spaces of densities}

There are some algebraic or analytic structures that are of great value and it is always profitabe to discuss them. 
We are pressed to ask if the set of all Lagrange densities of problem~(\ref{opeq1}) together 
with the zero function form a vector space.

\subsubsection{The vector spaces of Lagrange densities}

Let $\mathcal{V}$ denote the vector space of all functions of the form $F:\R^4\to\R$. We show that the 
set of all Lagrange densities $\mathcal{V}_\L$ of problem~(\ref{opeq1}) together with the zero function 
is a vector subspace of $\mathcal{V}$. The proof is actually trivial. 
If $\lambda_1,\,\lambda_2\in\R$ and $\L_1$, $\L_2$ 
are two Lagrange densities of problem~(\ref{opeq1}), then the Gateaux derivative of the action  
$$
    S[u] := \int_{I}\int_{\R^3} (\lambda_1\,\L_1[u] + \lambda_2\,\L_2[u])\,\d x \,\d t 
       =  \lambda_1\,S_1[u]  + \lambda_2\,S_2[u] 
$$
is given by 
$
  S'[u](h) = \lambda_1\,S_1'[u](h)  + \lambda_2\,S_2'[u](h) 
$ 
and thus the optimality conditions for $S_1[u]$ and $S_2[u]$ imply $S'[u](h)=0$ for $h\in\D$.  
But this means nothing else but $\lambda_1\,\L_1[u] + \lambda_2\,\L_2[u]\in\mathcal{V}_\L$, 
i.e. $\mathcal{V}_\L$ is a vector subspace of $\mathcal{V}$ and each element of it, except the zero vector, is  
a Lagrange density of problem~(\ref{opeq1}).  
If the Lagrange densities are absolute integrable, then $\mathcal{V}_\L$ is a normed vector space with the $L^1-$norm   
and it follows that $\mathcal{V}_\L$ is complete. We note that if $f$ is sufficiently nice, then the space 
$\mathcal{V}$ has more nice properties, but we shall not dwell on such questions. 

The crucial point is that there are many equivalent Lagrange densities for a given problem, some of them can easily be 
guesses and used to superpose more complex ones. In the language of gaming, it is an open world that want to be 
explored. 

\begin{rema}
The reader may think that  
$$
     \L_{triv} = \frac{1}{2}\,u^2 - (\oA^{-1}\,f)\,u  \qquad\mbox{and}\qquad  
     \tilde \L = -\frac{1}{2}\,u^2 
$$
must be equal, but this is wrong. The whole expression $\frac{1}{2}\,u^2 - (\oA^{-1}\,f)\,u$ is one vector in 
a vector space. Because $\frac{1}{2}\,u^2$ and $(\oA^{-1}\,f)\,u$ are not elements of this vector space, 
$\L_{triv}$ is not a sum of them. In particular, $\L_{triv}$ and $\tilde\L$ are not equal and not equivalent. 
We will see that the situation is different with Hamiltonians. The task of $\L$ is to model equations and 
Hamiltonians and nothing more. 
(What we permit is that $\L$ and $\tilde\L:=\L - f\,u$ are equal for $t>0$, where $f$ models initial 
conditions. Here $\L$ is the Lagrange density of a problem posed for all times and $\tilde\L$ is the Lagrange 
density of the equivalent initial valued problem posed only for positive time.)  
\end{rema}

\begin{rema}
We note that 
\begin{equation*}
   \L_{\oP_1 + \oP_2} \not =  \L_{\oP_1} + \L_{\oP_2}
   \qquad (\mbox{$\oP_1$, $\oP_2$ pdo's,})
\end{equation*}
due to non-linearity.
\end{rema}

\subsubsection{Vector spaces of Hamiltonian densities}

The \emph{Hamiltonian density} and its \emph{Hamiltonian} are usually defined by  
\begin{equation}\label{defHdensity}
   \H := (\partial_{u_t} \L) \, u_t - \L 
   \qquad\mbox{and}\qquad
   H := \int_{\R^3} \H\,\d \x\,,
\end{equation} 
respectively. This definition is only applicable if $\L$ depends at least on $u_t$.  
According to~(\ref{LP}), such a Lagrange density is for example given by
\begin{equation}\label{Lcomplex}
     \L
        = -\frac{1}{2}\,(\partial_\mu u)(\partial^\mu u) - (\partial_\mu\,\partial^\mu\,\oA^{-1}\,f)\,u\, 
\end{equation} 
and leads to 
\begin{equation}\label{Hcomplex}
\begin{aligned}
    \H = \frac{1}{2}\,\sum_{\mu=0}^3 (\partial_\mu u)^2 
          + (\partial_\mu\,\partial^\mu\,\oA^{-1}\,f)\,u \,. 
\end{aligned}
\end{equation} 
Or if we choose the Lagrange density $\tilde\L := \L_{\partial_t} = \frac{1}{2}\,u_t^2 + (\A^{-1} \,f_{tt})\,u$, 
we obtain the Hamiltonian density $\tilde \H = \frac{1}{2}\,u_t^2 - (\A^{-1} \,f_{tt})\,u$. 
Thus, if we apply the above definition of the Hamiltonian density to different (but equivalent) Lagrange densities, 
then we usually obtain different Hamiltonian densities. For some problems the first variant is conserved and for 
some other problems the second one (as shown later in some examples). 

We now show that the set of all these Hamiltonian densities with the zero function form a vector space. 
Let $\lambda_1,\,\lambda_2\in\R$, $\L_1,\,\L_2\in\mathcal{V}_\L$ depend (at least) on $u_t$ and $u$ ($f$ fixed), 
and let $\H_1$ and $\H_2$ denote the Hamiltonian densities of $\L_1$ and $\L_2$, respectively.
Then the Hamiltonian density of $\L := \lambda_1\,\L_1 + \lambda_2\,\L_2 \in\mathcal{V}_\L$ 
is given by 
$$
  \H 
      = (\partial_{u_t} [\lambda_1\,\L_1 + \lambda_2\,\L_2]) \, u_t - \lambda_1\,\L_1 - \lambda_2\,\L_2 
      = \H_1 + \H_2
$$ and thus the set of Hamiltonian densities together with the zero function is a vector subspace of $\mathcal{V}$.

As will be shown below, the definition~(\ref{defHdensity}) of an Hamiltonian density, is not the only one. 
Therefore, we could define the vector space of 'all' Hamiltonian densities, after defining all possible 
Hamiltonian densities. But we are not doing this in detail. Our point of view is that this space is a (large) 
vector subspace of $\mathcal{V}$, but with much more structure (not just vector addition and multiplication with 
a real skalar). It is reasonable to calculate with Hamiltonian in the same way as we calculate with $u$ or with  
equations, e.g. the Hamiltonian density $\H$ in~(\ref{Hcomplex}) is equal to  
$\frac{1}{2}\,\sum_{\mu=0}^3 (\partial_\mu u)^2 + (\partial_\mu\,\partial^\mu\,u)\,u$. 
Actually, this fact is nothing new, we just want to stress it to avoid confusion.

\subsubsection{Examples of Hamiltonian densities}

There are many different Hamiltonian densities out there. For example, for each Lagrange density of the form 
$\L[\oT\,u,\ldots,f]$,  where $\oT$ is an appropriate operator (possibly non-linear), there exists an 
Hamiltonian density  
\begin{equation}\label{defHT}
     \H  
           := \frac{ \partial \L}{ \partial(\oT u) }(\oT u) - \L \,.
\end{equation}

\begin{exam} 
From the trivial Lagrange density~(\ref{Ltrivial}), i.e. $\L_{triv}[u,f] = \frac{1}{2}\,u^2 - (\oA^{-1}\,f)\,u$, 
we get the trivial Hamiltonian density 
$$
  \H_{triv} := \H_{Id} = \frac{1}{2}\,u^2  \qquad\quad (\oT=\Id) \,.
$$ 
Conservation of the respective Hamiltonian $H_{triv}$ means that $\int_{\R^3} u^2\,\d \x$ is constant over time. 
For example, this is the case if $\oA = \partial_t + \v_0\cdot\nabla$ with $\v_0\in\R^3$ and 
$f(\x,t) = u_0(\x)\,\delta(t)$, i.e $u$ satisfies $u_t(\x,t) + \v_0\cdot(\nabla u)(\x,t) = u_0(\x)\,\delta(t)$.
\end{exam}

\begin{exam}\label{exam:massH}
Motivated by the previous examples, we suspect that there must be a Hamiltonian that describes total mass. 
Indeed, a Lagrange density of~(\ref{opeq1}) is given by 
$$
     \L_{M} := \sqrt{u}\,\sqrt{u} - 2\sqrt{ \oA^{-1}f }\,\sqrt{u} \qquad \mbox{(employs info $u\geq 0$)}
$$
and implies the required Hamiltonian density    
$$
   \H_M := \H_{ \sqrt{} }  = u  \qquad\quad\mbox{($\oT$ is non-linear)} \,. 
$$
Here we require a Lagrange density that is formulated as functions of $\sqrt{u}$, where $u$ denotes the 
density.  This Hamiltonian is conserved for all the mass transport equations endowed with initial conditions.  
\end{exam}

If $\oT$ is linear with the adjoint $\oT^*$, then there exists a \emph{Lagrange density} with a respective \emph{Hamiltonian density}, namely  
\begin{equation}\label{defLT}
       \L_\oT := \frac{1}{2}\,(\oT\,u)^2 - (\oA^{-1}\oT^*\oT\,f)\,u  
          \quad\mbox{and}\quad 
       \H_{\oT} := \frac{1}{2}\,(\oT\,u)^2 + (\oA^{-1}\oT^*\oT\,f)\,u \,.
\end{equation}
If $\oT$ is design in the right way for given $\oA$ (and $f$ models initial conditions), then this Hamiltonian is conserved. 

Let us inspect the following case  
\begin{equation}\label{defLdt}
   \L_{\partial_t} = \frac{1}{2}\,u_t^2 + (\partial_t^2 \oA^{-1}\,f)\,u  \qquad\mbox{and}\qquad 
   \H_{\partial_t} = \frac{1}{2}\,u_t^2 - (\partial_t^2 \oA^{-1}\,f)\,u
\end{equation}
in more detail. 
As usual, let $u$ solve $\oA\,u=f$ and let $f$ be a finite sum of products of the form 
$\phi_j(\x)\,\delta^{(j)}(t)$. If $\oQ := \partial_t^2 \oA^{-1}$ is a pdo, then $\oQ\,f$ vanishes for positive time 
and thus we infer for positive time 
$$
  \frac{\d H_{\partial_t} }{\d t} 
     = \int_{\R^3} \left[  u_t\,(u_{tt} - \oA^{-1}\,f_{tt}) - (\oQ\,f_t)\,u \right]  \, \d \x 
     = - \int_{\R^3} (\oQ\,f_t)\,u   \, \d \x = 0 \,.    
$$
That is to say, the Hamiltonian $H_{\partial_t} = \frac{1}{2}\int_{\R^3} u_t^2  \, \d \x$ is conserved.  
Let us consider the following concrete problem 
$$
    u_{tt} (x,t) = \phi_0(x)\,\delta'(t) +  \phi_1(x)\,\delta(t)  \qquad\quad (\oA = \partial_t^2) \,,
$$ 
where each $\phi_j$ is a nice function of $x$. Then we have $\oQ := \partial_t^2 \oA^{-1} = \Id$ and thus the quantiy 
$\frac{1}{2}\int_{\R^3} u_t^2  \, \d \x$ is conserved. It is clear that this Hamiltonian is also 
conserved if $\oA = \partial_t$ holds, but not if $\oA = \partial_t^n$ with $n\geq 3$. 
If we consider the same problem but the Lagrange density induced by $\oT := \partial_t \partial_x$, then it follows that 
$\oQ := \partial_t^2 \partial_x^2\oA^{-1} = \partial_x^2$ and 
$$
  \frac{\d H_{\partial_{tx}} }{\d t} 
     = \int_{\R^3} \left[  u_t\,\partial_t^2\partial_x^2 (-u+\oA^{-1}\,f) + f_{txx}\,u \right]  \, \d \x 
     = \int_{\R^3} f_{txx}\,u   \, \d \x = 0 \,.    
$$
Hence $H_{\partial_{tx}} = \frac{1}{2}\int_{\R^3} u_{tx}^2  \, \d \x$ is conserved, too. 
In summary, the Lagrange densities given by~(\ref{defLT}) with $\oT=\partial_t$ and $\oT=\partial_t\partial_x$ tell us for a given problem 
whether there are (special) conserved quantities or not. In this example, at least $H_{\partial_t}$ and $H_{\partial_{tx}}$ 
are conserved.  

Of course, the situation is more complicated for other equivalent Lagrange densities. The crucial point is that 
each Lagrange density of a problem can tell us something, we just need to understand it and look for the most 
'illuminating' ones.\\

\begin{exam} 
As shown above, $u^2$ is an Hamiltonian density. We can also consider $u\,u^*$ instead of $u^2$, where $u^*$ denotes 
the conjugate complex of the (now) complex valued $u$ (e.g. probability amplitude). Because 
\begin{equation*}
     \L_P[u,u^*,f] := \frac{1}{2}\,u\,u^* - \left[ (\oA^{-1}\,f)\,u^* + (\oA^{-1}\,f)^*\,u\right]  \,.
\end{equation*} 
is a Lagrange density of~(\ref{opeq1}), because the optimality condition reads as follows 
$$
  0 = \int_{\R^3} \left[ (u-\oA^{-1}\,f)\,h^* + (u^*-(\oA^{-1}\,f)^*)\,h \right] \,\d \x\,
$$
for all complex-valued functions $h$ with real and imaginary part out of $\D$. 
The Hamiltonian density defined by  
\begin{equation*}
\begin{aligned}
   \H_P
          := \frac{1}{2}\,\left[ (\partial_u \L_P)\,u + (\partial_{u^*} \L_P)\,u^* \right]  - \L_P \,. 
\end{aligned}
\end{equation*} 
leads to $\H_P = \frac{1}{2}\,u\,u^*$. If $u$ denotes a probability amplitude and $H_P$ is conserved, then the 
total probability is conserved. Of course, the Schr\"odinger equation is such an example. We note 
that problems with higher order Lagrange equations can be formulated as a system of first order equations and 
for both there are equivalent Hamiltonian densities. 
\end{exam}

\begin{rema}
The usual Lagrange density $\L$ and Hamiltonian density $\H$ of the wave equation $-\partial_\mu\,\partial^\mu\,u = f$ 
are given by $\L = -\frac{1}{2}\,(\partial_\mu\,u)\, (\partial^\mu\,u) + f\,u$ and 
$\H = \frac{1}{2}\,\sum_{\mu=0}^3(\partial_\mu\,u)^2 - f\,u$ with $c=1$. (Cf.~(\ref{Lcomplex}) and~(\ref{Hcomplex}) 
with $\oA = -\partial_\mu\,\partial^\mu$.) Similarly as above, it follows that 
$$
  \frac{\d H }{\d t} 
     = \int_{\R^3} \left[  u_t\,(-\partial_\mu\partial^\mu \,u - f) - f_t\,u \right]  \, \d \x 
     = - \int_{\R^3} f_t\,u   \, \d \x = 0 \,,    
$$
if $f$ models initial conditions. We note that the Lagrange density does not depend explicit on time if $f$ models 
initial conditions. (Indeed, it can be proven that the converse is also true. Cf.~Subsection~\ref{subsec-note}.)   
\end{rema}

\subsubsection{Higher order Hamiltonians}

Let $H$ be a conserved Hamiltonian with density $\H$ for the problem $\oA\,u=f$, where $f$ models initial conditions. 
Let us construct 'higher order' Hamiltonian densities from $\H$. 
It is clear that $\tilde u := \partial_t^n u$ ($n\in\N$) satisfies $\oA\,\tilde u = \partial_t^n \,f =: \tilde f$, where  
$\tilde f$ models also initial conditions. And thus $\H[\tilde u]$ is also conserved. This leads us to the following 
definition of higher order Hamiltonian densities  
\begin{equation}\label{defHN}
     \H^n[u,f] := \H [\partial_t^{n-1} u,\partial_t^{n-1} f] \qquad (n\in\N)\,.   
\end{equation}

Let us start with a simple example.

\begin{exam}
Let us consider the Lagrange density~(\ref{defLdt}), i.e. 
$\L_{\partial_t} = \frac{1}{2}\,u_t^2 + (\partial_t^2 \oA^{-1}\,f)\,u$, which implies 
$\H_{\partial_t} = \frac{1}{2}\,u_t^2 - (\partial_t^2 \oA^{-1}\,f)\,u$. Then we have the following higher order 
Hamiltonian densities: 
\begin{equation*}
   \H_{\partial_t}^n = \frac{1}{2}\,(\partial_t^n u)^2 - (\oA^{-1}\partial_t^{n+1} \,f)\,(\partial_t^{n-1} u) 
   \qquad\quad (n\in\N)\,. 
\end{equation*}
Let us compare this Hamiltonian with $H_{\partial_t^n}$, which reads as follows 
$$
    \H_{\partial_t^n}
       := \frac{\partial \L_{\partial_t^n}}{\partial (\partial_t^n u)}\,(\partial_t^n u) - \L_{\partial_t^n} 
        = \frac{1}{2}\,(\partial_t^n u)^2 + (-1)^n (\partial_t^{2 n} \oA^{-1}\,f)\,u \,,
$$
due to $\L_{\partial_t^n} = \frac{1}{2}\,(\partial_t^n u)^2 - (-1)^n (\partial_t^{2 n} \oA^{-1}\,f)\,u$. 
For this class of problems, $\H_{\partial_t^n}$ is conserved if $\H_{\partial_t}$ is conserved. Strictly speaking 
$\H_{\partial_t}^n \not= \H_{\partial_t^n}$, but if $f$ models initial data, then $\H_{\partial_t}^n = \H_{\partial_t^n}$ 
for $t>0$.  
\end{exam}

\begin{exam}\label{exam:Adv}
Let us consider the following advection problem  
$$
    u_t + \v_0\cdot(\nabla\,u) = u_0\,\delta =: f \quad\mbox{with}\quad u|_{t<0} = 0 
$$ 
For convenience, we use the notation $\oA := \partial_t + \oB$ and $\oB\,u := \v_0\cdot(\nabla\,u)$. 
For this example, we have $\oB^* = -\oB$. If we apply to this equation the time reversal 
$\oA^\# := -\partial_t + \oB$, then we obtain the equivalent equation 
$$
   u_{tt}(\x,t) - (\oB^2\,u)(\x,t) = -(\oA^\# f)(\x,t) = u_0(\x)\,\delta'(t) - (\oB u_0)(\x)\,\delta(t) \,. 
$$
From this we infer the following Lagrange density and Hamiltonian density 
$$
    \L = \frac{1}{2}\,u_t - \frac{1}{2}\,(\oB\,u)^2 - (\oA^\#\,f)\,u  \quad\mbox{and}\quad
    \H = \frac{1}{2}\,u_t^2 + \frac{1}{2}\,(\oB\,u)^2 + (\oA^\#\,f)\,u\,.
$$
Because $\oA^\#\,f$ vanishes for positive time, the respective Hamiltonian is conserved, more precisely 
$$
  \frac{\d H}{\d t}[u] 
       = \int_{\R^3}  \left[ u_t\,\left( u_{tt} - \oB^2(u) + \oA^\#\,f \right) 
                        + (\oA^\#\,f_t)\,u\right] \,\d \x 
       = 0  \qquad\mbox{for}\qquad t>0\,.
$$
As a consequence, the respective higher order Hamiltonians 
$H^n = \int_{\R^3}  (\oB\,\partial_t^n u)^2  \,\d \x = \int_{\R^3}  (\oB^{n+1} u)^2  \,\d \x$ 
$(n\in\N)$ are also conserved for $t>0$. (Here we have used that $u_t = -\oB\,u$ for $t>0$.) Because the mass 
is conserved for advection, it follows that each Hamiltonian 
$H_M^n :=\int_{\R^3}  \partial_t^n u  \,\d \x = (-1)^n \int_{\R^3}  \oB^n u  \,\d \x$ 
is also conserved. These conservation laws are not obvious or trivial. 
\end{exam}

\section{Two case examples and a note}
\label{sec-CaseEx}

Let us make a break and apply the previous results 
\begin{itemize}
\item  to Airy's equation and 
\item  to the standard frequency dissipative wave equation. 
\end{itemize}
Moreover, we have to make a note about special source terms and initial conditions.

\subsection{Airy's equation}

The $1D-$\emph{Airy's equation} reads as follows 
$$
        u_t + u_{xxx} = \varphi \,\delta =: f  \qquad\quad (\oA = \partial_t + \partial_t^3) \,,
$$
where $\varphi$ vanishes for sufficiently large $|x|$. Let us have a closer look. 
As shown previously, the Lagrange density $\L_M = u - 2\,\sqrt{A^{-1}\,f}\,\sqrt{u}$ implies the Mass-Hamiltonian density 
$\H_M = u$, which is conserved. Indeed, we have for $t>0$: 
\begin{equation*}
\begin{aligned}
   \H'_M (t)
       &= \int_{\R^3} u_t(x,t) \,\d x 
       = -\int_{\R^3} u_{xxx}(x,t)\, \d x 
       = \lim_{z\to \infty } [u_{xx}(-z,t) - u_{xx}(z,t)] \\   
       &= \lim_{z\to \infty } [\varphi_{xx}(-z) - \varphi_{xx}(z)]  
       = 0 \,. 
\end{aligned}
\end{equation*}
As a consequence, Airy's equation describes a mass transport problem.

If we apply the time reversal operator $\oA^\# := -\partial_t + \partial_x^3 $ to the original problem, 
we obtain the equivalent equation $\partial_t^2 u - \partial_x^6 u = \oA^\#\,f$, which leads to the Lagrange density 
$$
     \L = \frac{1}{2}\,u_t^2 - \frac{1}{2}\,u_{xxx}^2 + (\oA^\#\,f)\,u \,.
$$
This $\L$ and its respective Lagrange equation only contain partial differential operators and as a consequence, 
if $f$ models initial data, then $\oA^\#\,f$ models initial data, too. Hence we can pose the problem on the positive 
time line and use 
$$
     \L = \frac{1}{2}\,u_t^2 - \frac{1}{2}\,u_{xxx}^2  \quad\mbox{for}\quad t>0\,,
$$
and endow the respective (second order) Lagrange equation with the initial conditions $u(\cdot,0+)=\varphi$ and 
$u_t(\cdot,0+)=-\varphi_{xxx}$. (The last condition follows from $u_t = -u_{xxx}$ for $t>0$.) This Lagrange density 
implies the following Hamiltonian density 
$$
    \H := (\partial_{u_t}\L) \,u_t - \L = \frac{1}{2}\,u_t^2 + \frac{1}{2}\,u_{xxx}^2  \quad\mbox{for}\quad t>0\,
$$
that is conserved, due to
\begin{equation*}
\begin{aligned}
   \H' (t)
       &= \int_{\R^3} [u_t\,u_{tt} + u_{xxx}\,u_{xxxt}] \,\d x 
       = \int_{\R^3} u_t\,[\partial_x^6 u - \partial_x^6 u] \,\d x 
       = 0 \,.
\end{aligned}
\end{equation*}
This Hamiltonian density is actually nothing else but $\tilde \H = u_t^2$, due to $u_t = -u_{xxx}$ for $t>0$. 
In this example, $H$ is not the wave energy. 
There are much more conserved quantities (if the function $u$ is sufficiently smooth). For example, 
$$
    \H_M^n[u] = \int_{\R^3} \partial_x^{3 n} u\,\d x = 0 
\qquad\mbox{and}\qquad  
   \H^n[u] = \int_{\R^3} \left[ \partial_x^{3 n} u \right]^2\,\d x = 0  
$$ 
for $t>0$ and $n\in\N$. 
It is not obvious that $u$ satisfies these conditions, due to its spatial properties. $u$ satisfies these properties, 
because it solves Airy's equation. \\

This example shows in an impressive way, that there are many different Hamiltonians that can be conserved 
and which give us vital information.  
If we consider a dissipative system, then some (or all?) Hamiltonians are not conserved but nevertheless 
they provide us with important information. It is just much more difficult to extract the information. 
As far as we know,  it is not common to use the principle of least action for dissipative problems.

\subsection{Frequency dependent dissipative waves}

In this subsection, we want to demonstrate that the principle of least action is also applicable to (frequency dependent) 
dissipative waves. (Cf. also Subsection~\ref{subsec-FracExam}.)
The standard equation for such waves is an integro-differential equation of the form 
\begin{equation}\label{StandDissEq}
   \oA_\lambda\,u=f  \quad\mbox{with}\quad
   \oA_\lambda\,u :=
    \left( \frac{1}{c}\,\partial_t + \lambda\,D_* \right)^2 \,u 
      - \rho\,\nabla\cdot\left( \frac{\nabla u}{\rho} \right) \,,
\end{equation}
where $\lambda\geq 0$, $D_*$ denotes the \emph{time convolution operator} defined by 
$\F \{D_*\,u\} := \alpha_*\,\F \{u\}$ ($\F$ Fourier transform) and $\Re(\alpha_*)$ denotes the 
\emph{attenuation law} (cf. e.g.~\cite{KoScBo10,KoSc12}). 
The Telegraph equation and the Nachman, Smith and Waag equation are also included in this equation (as special cases 
without memory). 
It is best to discuss the Telegraph equation first. It reads as follows 
\begin{equation}\label{StandTelEq}
   \frac{1}{c_0^2}\,u_{tt} + \frac{d_0}{c_0}\,u_t - \Delta\,u = f  
   \qquad\quad \mbox{($c_0=1$, $d_0$ constants),}
\end{equation}
where $f(\x,t) := \varphi(\x)\,\delta'(t) + \psi(\x)\,\delta(t)$ models the initial conditions $u(x,0+) = \varphi(x)$ 
and $u_t(x,0+) = \psi(x)$. Then $\oA = -(\partial_\mu\partial^\mu + d_0\,\partial^0)$ and according 
to~(\ref{Lcomplex}) a Lagrange density is given by 
\begin{equation*}
     \L 
        = -\frac{1}{2}\,(\partial_\mu u)(\partial^\mu u) - (\partial_\mu\,\partial^\mu\,\oA^{-1}\,f)\,u\,. 
\end{equation*}
Here $\partial_\mu\,\partial^\mu\,\oA^{-1}$ is linear but not a partial differential operator (pdo).   
Hence the usual Hamiltonian density $\H:=\frac{\partial \L}{\partial u_t} - \L$ can be written as 
\begin{equation*}
\begin{aligned}
    \H = \frac{1}{2}\,\sum_{\mu=0}^3 (\partial_\mu u)^2 
          + (\partial_\mu\,\partial^\mu\,\oA^{-1}\,f)\,u \,. 
\end{aligned}
\end{equation*} 
At this point, it is reasonable to introduce the Energy-Hamiltonian density by 
$$
  \H_E 
      := \frac{1}{2}\,\sum_{\mu=0}^3 (\partial_\mu u)^2  
      = \H -  (\partial_\mu\,\partial^\mu\,\oA^{-1}\,f)\,u \,,
$$ 
which is actually the wave energy.  
From this together with $u - \oA^{-1}\,f = 0$ and $u_t = \oA^{-1}\,f_t$, we infer 
\begin{equation*}
\begin{aligned}
     \frac{\d H}{\d t} 
        &= \int_{\R^3} \left\{ (\partial^0\,u)\,(\partial_\mu \partial^\mu\,
                                               [ u - \oA^{-1}\,f])       
                         - (\partial_\mu\,\partial^\mu\,\oA^{-1}\,\partial^0\,f)\,u 
                              \right\} \,\d \x \\
        &= \int_{\R^3} (\partial_\mu\,\partial^\mu\,\oA^{-1}\,\,\partial_0f)\,u  \,\d \x                                                      
\end{aligned}
\end{equation*}
and thus 
\begin{equation}\label{changeH}
\begin{aligned}
     \frac{\d H_E}{\d t} 
        &= - \int_{\R^3} (\partial_\mu\,\partial^\mu\,\oA^{-1}\,f)\,(\partial_0\,u)  \,\d \x                              
        = -\int_{\R^3}  (\partial_\mu\,\partial^\mu\,u)\,(\partial_0\,u) \,\d \x   \,.                        
\end{aligned}
\end{equation}
For the Telegraph equation this simplifies to 
$\frac{\d H_E}{\d t} = - \int_{\Omega} d\,(\partial_0 u)^2 \,\d \x \,\d t \leq 0\,$\,, which is well-known. 
Because $H_E$ is positive and conserved, it can be interpreted (up to a constant) as the energy of the 
considered dissipative system. 

The above calculations that leads to~(\ref{changeH}) can equally be performed for the dissipative wave 
equation~(\ref{StandDissEq}). Analogously, it follows that a Lagrange density 
$\L_\lambda$ and a Hamiltonian density $\H_\lambda$ are given by~(\ref{Lcomplex}) and~(\ref{Hcomplex}), 
respectively, if $\oA$ is replaced by $\oA_\lambda$. Moreover, the change of the Hamiltonian (energy) is given by 
$\frac{\d H_\lambda}{\d t} = -\int_{\R^3}  (\partial_\mu\,\partial^\mu\,u)\,(\partial^0\,u) \,\d \x$\,,  
which must be non-positive and thus implies a condition on the operator $\D_*$. 
Of course, this matter is very complicated and it needs considerable afford to get the necessary understanding 
(with and without the principle of least action). At any rate, this demonstrates the usefulness of the principle of 
least action to waves including frequency dependent dissipative waves. 

\begin{rema}
In Subsection~\ref{subsec-FracExam}, we show that a Lagrange density of the fractional diffusion equation 
$(\partial_t^{ \frac{1}{2} } u)(\x,t) + (\oB \,u)(\x,t) = \varphi(\x)\,(\partial_t^{ -\frac{1}{2} } \delta)(t)$  
with $\oB := \nabla \cdot(\v \quad)$ is given by 
\begin{equation*}
\begin{aligned}       
   \L(u) 
      = \frac{1}{2} (\partial_t u)^2 + \frac{1}{2} (\oB^2 u)^2 
         + u\,(\partial_t + \oB^2) f_1  
         + g_0 \, u \,,
\end{aligned}
\end{equation*}
where $ f_1 = - (\oB \varphi) \, (\partial_t^{ -\frac{1}{2} } \delta) $ and 
$g_0 = \varphi\,\delta' + (\oB^2 \varphi)\,\delta$. Here $\varphi$ is nothing else but the initial valued 
$u(\cdot,0+)$. It will be shown that the Mass-Hamiltonian is conserved for this fractional diffusion problem. 
\end{rema}

\subsection{Note about initial conditions}
\label{subsec-note}

It is well-known that initial conditions of the $1D-$standard wave equation, i.e. $u(x,0+) = \varphi(x)$ and 
$u_t(x,0+)= \psi(x)$ can be modeled by the source term $f(\cdot,t) = \varphi\,\delta'(t) + \psi\,\delta(t)$. 
But what happens if the source term is not of this form, say $2\,\phi(x)\,\delta''(t)$. Is the energy still 
conserved? And does this source term describe somehow initial data? We show that alternatively the source term 
$\tilde f(x,t)=2\,\phi_{xx}(x)\,\delta(t)$ can be used, which models the initial data $u(\cdot,0+) = 0$ 
and $u_t(\cdot,0+)= \phi_{xx}$. 
The presented simple trick can be use to show that each source term of the form $f(x,t) = \sum_{j=1}^n \phi_j(x)\,\delta^{(j)}(t)$  models initial data for any pde. \\

Let us consider the two problems 
$$
   v_{tt}(x,t) - v_{xx}(x,t) = \phi(x)\,\delta''(t) =: f(x,t)  
$$
and
$$
   \tilde v_{tt}(x,t) - \tilde v_{xx}(x,t) = \phi_{xx}(x)\,\delta(t) =: \tilde f(x,t) 
$$
for $(x,t)\in\R^2$, where $\phi$ is $2$-times differentiable. Here $\phi$ is not an initial data for the wave $v$ and 
$\phi_{xx}$ is an initial data for the wave $\tilde v$. If $w$ solves the problem 
$$
   w_{tt}(x,t) - w_{xx}(x,t) = \phi(x)\,\delta(t) 
$$
for $(x,t)\in\R^2$, then we have $w_{tt} = v$ and $w_{xx} = \tilde v$. Because of $w_{tt} = w_{xx}$ for 
positive time, we infer $v = \tilde v$ for each time $\not=0$. Therefore the source term $f$ can be replaced 
by the source term $\tilde f$ that models the initial conditions 
$$
    v(\cdot,0+)=0    \quad\mbox{and}\quad   v_t(\cdot,0+)=\phi_{xx}\,.
$$
So after all, the source term can be formulated, via a trick, as initial data. 
It is clear that the same idea can be applied to any pde with a source term that is a finite sum of terms like  
$\phi(\x)\,\delta^{(m)}(t)$ ($m\in\N$).

\section{Special Lagrange densities}
\label{sec-pdoL}

So far we have seen that there are many Lagrange densities. Actually, there exists not the best Lagrange density, 
each of them has its purpose. In the long run, we have to inspect all of then and extract the vital information out 
of them such that it is ready to be used. (This is not done in this paper.) Let us do part of this task now. We look 
for Lagrange densities that contain only partial differential operators (pdo's). We call them pdo-Lagrange 
densities. \\

\noindent 
If $\oA$ is a linear partial operator, then (in general) the Lagrange density 
$$
    \mbox{ $\L= -\frac{1}{2}\,\partial_\mu\partial^\mu u - (\partial_\mu\partial^\mu\,\oA^{-1}\,f)\,u$ 
           \quad 'hides' some \emph{information} }
$$
about the considered problem in its last term. More precisely, the last term may not be a pdo. 
For example, if we consider the Telegraph equation, then $\partial_\mu\partial^\mu\,\oA^{-1}$ is not a pdo 
($\oA = \partial_\mu\partial^\mu + d\,\partial_0$). 
$$
\mbox{ The same is true for \quad $-\partial_\mu\partial^\mu u = -\partial_\mu\partial^\mu\,\oA^{-1}\,f$, }
$$
its respective Lagrange equation. Naturally, we would like to extract this information, i.e. to write 
it as a pdo. According to~(\ref{LP}), i.e. $\L_\oP[u,f] := \frac{1}{2}\,(\oP\,u)^2 - (\oP^*\,\oP\,\A^{-1} \,f)\,u$, 
we require a linear (injective) pdo $\oP$ such that $\oP^*\,\oP\,\A^{-1}$ is  
a pdo, say $\oQ$. Then the respective Lagrange equation is given by the pde 
\begin{equation}\label{FormQ}
   \oP^*\,\oP\,u = \oQ\,f  \qquad\mbox{,where}\qquad 
   \L := \frac{1}{2}\,(\oP\,u)^2 - (\oQ \,f)\,u\,.
\end{equation}
If $\oA$ is a linear pdo, then we can always choose $\oP := \oA$, which leads to 
$\L_{\oA} := \frac{1}{2}\,(\oA\,u)^2 - (\oA^*\,f)\,u$ and thus the respective Lagrange equation is  
\begin{equation*}
   \oA^*\,\oA\,u = \oA^*\,f  \qquad\quad\mbox{\emph{the normal equation}}\,.
\end{equation*}
We recall that if $\oA$ is injective, then the normal equation has the same unique solution as the original problem. 
(We can equally choose $\oP:=\oA^*$ (if $\oA$ and $\oA^*$ commute), which leads to the Lagrange density 
$\L_{\oA^*} := \frac{1}{2}\,(\oA^*\,u)^2 - (\oA^*\,f)\,u$.)  
This choice suggests the following Hamiltonian density 
$$
    \H_\oA 
        = (\partial_{\oA\,u} \L)\,(\oA\,u) - \L 
         = \frac{1}{2}\,(\oA\,u)^2 + (\oA^*\,f)\,u 
         = \frac{1}{2}\,f^2          + (\oA^*\,f)\,u \,,
$$
which vanishes for $t>0$ if $f$ models initial data. 
In other words, this Hamiltonian is conserved but because it is the zero function it cannot be exploited 
(cf. also Example~\ref{exam:AdvDiff}). Thus the normal equation (of a given problem) leads alway to a 
pdo-Lagrange density, however, its respective Hamiltonian $H_\oA$ is just the zero function. (Of course, 
other  non-trivial Hamiltonians may be conserved.)

\begin{rema}
A main advantage of the formulations in~(\ref{FormQ}) is that if $f$ models initial data, then $\oQ f$ 
models initial data, too. (In general, this is not true, if $\oQ$ is an integral operator like a time-fractional 
operator or the inverse of a partial differential operator.) In particular, then $f$ as well as $\oQ f$ vanish for 
positive time. This is very useful to find and investigate conserved Hamiltonians. This is an advantage of 
pdo-Lagrange densities. \\
\end{rema}

Let us consider a large class of problems that contains advection as well as diffusion.  

\begin{exam}\label{exam:AdvDiff}
Let $\oB$ be a linear partial differential operator in $\x$ such that $\oB^* = s_0\,\oB$ with $|s_0|=1$ and 
$\oB^*$ denote its adjoint w.r.t. $\x$.\footnote{Here we tacitly assume that $\oB$ is not of the form $u_t + \oB_1$. 
This example remains true if $\oB$  contain operators of the form $\partial_t^{2 m}$ for $m\in\N$.} 
We consider the problem 
$$
    u_t + \oB\,u = u_0\,\delta =: f \quad\mbox{with}\quad u|_{t<0} = 0 
    \qquad\quad (\oA := \partial_t + \oB)\,. 
$$ 
Here $f$ models the initial data $u(\x,0+) = u_0(\x)$. If we apply to this equation the time reversal 
$\oA_{tr} := -\partial_t + \oB$ or its adjoint $\oA^* := -\partial_t + \oB^*$, then we obtain  
the equivalent equations 
$$
    u_{tt} - \oB^2\,u = -\oA_{tr} f  \qquad\mbox{and}\qquad 
    u_{tt} + (1-s_0)\,\oB\,u_t - s_0\,\oB^2\,u = -\oA^* f   
$$
that have the Lagrange densities 
$$
    \L_1 = \frac{1}{2}\,u_t^2 + \frac{1}{2}\,s_0\,(\oB\,u)^2 - (\oA_{tr} f)\,u  
                  \qquad\mbox{and}\qquad 
    \L_2 = \frac{1}{2}\,(u_t + \oB\,u)^2 - (\oA^*\,f)\,u  \,,
$$
respectively. 
Because $\oA_{tr} f$ and $\oA^*\,f$ vanish for positive time, the respective Hamiltonian densities of 
these two Lagrange densities read as follows  
$$
    \H_1 = \frac{1}{2}\,u_t^2 - \frac{1}{2}\,s_0\,(\oB\,u)^2  
                  \qquad\mbox{and}\qquad 
    \H_2 = \frac{1}{2}\,u_t^2 - \frac{1}{2}\,(\oB\,u)^2  
    \qquad\mbox{for}\qquad t>0 \,,
$$
which are equal if $\oB$ is self-adjoint ($s_0=1$). 
In case of advection with constant velocity $\v_0$ $(\oB = \v_0\cdot\nabla)$, the Hamiltonian $H_1$ is the 
conserved energy and the other Hamiltonian $H_2$ is just the zero function ($=$ Hamiltonian of normal equation). 
In case of diffusion, i.e. $(\oB = -\nabla \cdot (D\nabla \quad ))$, both Hamiltonians are equal to 
the zero function.
\end{exam}

\begin{rema}
1) That the standard model of diffusion does not have a reasonable energy is not entirely surprising. 
It is well known that the solution of the diffusion equation for an initial mass (concentrated in a Ball) is 
positive everywhere for an arbitray small (finite) positive time. So some particles must propagate arbitrarily fast. 
How can such a model process have a reasonable 
finite energy? \\
2) How can the existence of different Hamiltonians for equivalent Lagrange densities be explained? First of all, 
in general, the considered (Lagrange) equations describe different processes if general source terms are used and 
not (our) special ones. Because the Hamiltonians are non-linear, they are usually not equal for special  
source terms. 
\end{rema}

Let us shortly discuss how a reasonable energy may be defined? This is a generalisation of the usual Hamiltonian. 
Let us assume that the problem can be reformulated such that it reads as follows 
$$
   \oA\,u = \oQ\,f  \quad \mbox{,where}\quad \oA = \sum_{j=1}^n s_j\,\oA_j^*\,\oA_j 
   \quad\mbox{with}\quad 
   \oA_j^* = -\oA_j \quad\mbox{and}\quad  |s_j|=1\,. 
$$
As before, $\oQ$ denotes a pdo and $f$ models initial data. Without loss of generality, we assume that $s_j=+1$ for 
$j\in\{1,\,2,\,\ldots,\,m\}$ and $s_j=-1$ for $j\in\{m+1,\,m+2,\,\ldots,\,n\}$. For this problem, a Lagrange density 
is given by 
$$
  \L = \frac{1}{2}\,\sum_{j=1}^n s_j\,(\oA\,u)^2 + (\oQ\,f)\,u \,.
$$
Then the following Hamiltonian density is alway non-negative for $t>0$ 
$$
   \H_+ 
     := \sum_{j=1}^m \frac{\partial \L}{\partial (\oA_j\,u)} (\oA_j\,u) - \L 
      = \frac{1}{2}\,\sum_{j=1}^n (\oA\,u)^2 - (\oQ\,f)\,u\,.
$$
Similarly, the Hamiltonian density for $t>0$  
$$
   \H_- 
     := \sum_{j=m+1}^n \frac{\partial \L}{\partial (\oA_j\,u)} (\oA_j\,u) - \L 
      = - \frac{1}{2}\,\sum_{j=1}^n (\oA\,u)^2 - (\oQ\,f)\,u\,
$$
is always non-positive. 
Moreover, we have $\H_+ + \H_- = - 2\,(\oQ\,f)\,u$, which vanishes for positive time. In case of the standard 
wave equation, we have $m=1$, $n=3$ and $\oA_j = \partial_j$.

\section{Covariant Lagrange densities}
\label{sec-covL}

A main advantage of Lagrange densities is that if they are covariant, then each of them implies  
a covariant Lagrange equation (cf.~\cite{SuHr13,SuFr20a,SuFr20b}). Of course, we have already found 
a covariant Lagrange density in the trivial Lagrange density $\L_{triv}= \frac{1}{2}\,u^2 - (\oA^{-1}\,f)\,u$ 
introduced by~(\ref{Ltrivial}). However, it is usually not a pdo-Lagrange density. 
Let us consider some examples.

\begin{exam}\label{exam:TelEq}
With the operator $\oA := -(\partial_\mu\partial^\mu + d_0\,\partial^0)$ the Telegraph equation (for a homogeneous medium) 
can be written as $\oA\,u=f$. We assume that $f$ models the initial conditions $u|_{t=0} = \varphi$ and 
$u_t|_{t=0} = \psi$. Its equivalent  normal equation   
\begin{equation}\label{}
    [(\partial_\nu\partial^\nu) \,(\partial_\mu\partial^\mu) + d_0^2\,(\partial_0\partial^0)]\,u = \oA^*\,f\, 
\end{equation}
implies the following Lagrange density 
\begin{equation}\label{normalTel}
    \L
         = \frac{1}{2}\,(\partial_\mu\partial^\mu\,u)^2  
               + \frac{1}{2}\,d_0^2\,(\partial_0\,u)^2 
               - (\oA^*\,f)\,u\, 
\end{equation} 
and vice versa. 
The relativistic variant of the above Lagrange density can be readily obtained if 
$\frac{1}{2}\,d_0^2\,(\partial_0\,u)^2$ is replaced by $\frac{1}{2}\,d_0^2 \left( \frac{\d u}{\d \tau} \right)^2$. 

The source term $(\oA^*\,f)\,u$ in $\L$ can be removed if the Lagrange equation is endowed with the following 
initial conditions:
$$
     u|_{t=0} = \varphi   \,,\quad
     u_t|_{t=0} = \psi \,,\quad 
     u_{tt}|_{t=0} = - d_0\,c_0\,\psi + c_0^2\,\Delta\,\varphi   
$$
and 
$$
    u_{ttt}|_{t=0} 
        =  d_0^2\,c_0^2\,\psi 
            - d_0\,c_0^3\,\Delta\,\varphi  
            + c_0^2\,\Delta\,\psi   \,,
$$
due to $u_{tt} = - d_0\,c_0\,u_t + c_0^2\,\Delta\,u$.  
We see that the data $\varphi$ and $\psi$ together with the constants $c_0$ and $d_0$ are sufficient 
for the additional higher order initial conditions, which are readily derived from the original equation. 
\end{exam}

\begin{rema}
Do we need higher order equations? We ask this question independently from the above considerations. 
Due to their complexity, we naturally want to avoid them. However, it is very  
likely that some of them will play a prominent role in the future. For example, the higher order wave 
equation from Nachman, Smith and Waag (cf.~\cite{NacSmiWaa90}) has already found its way into 
Photoacoustic imaging (PAT) and we think it will play a prominent role there. \\
\end{rema}

Let us derive a covariant form of Airy's equation in $2D-$space-time (endowed with initial conditions).

\begin{exam}
The Airy equation reads as follows 
$$
   \oA\,u =0 \quad\mbox{with}\quad u|_{t=0} = \varphi 
      \quad\mbox{,where}\quad 
  \oA := \partial_0 + \partial_1^3\,.
$$ 
According to~(\ref{LP}) we require 
a partial differential operator $\oP$ such that (i) $\oQ := \oP^*\,\oP\,\A^{-1}$ is a partial differential operator and 
(ii) $\oP^*\,\oP$ is covariant, because then 
\begin{equation*}
   \mbox{ $\L_\oP := \frac{1}{2}\,(\oP\,u)^2 - (\oQ \,f)\,u$  \quad serves our purpose. }   
\end{equation*}
Note $\A^{-1}\,f$ is invariant and thus $\oQ\,f = \oP^*\,\oP\,\A^{-1}\,f$ is invariant, too. 
An operator $\oP$ can also be derived by sharp thinking, it follows 
$$
   \oA 
     = \partial_0 + \partial_1^3 
     = \partial_0 - \partial_0\partial^0 + \partial_0(\partial_\mu\partial^\mu)  
          + v_\nu\partial^\nu (\partial_\mu\partial^\mu) =: \oP 
          \qquad (v_0 := v_1 := 1).
$$
We note that $\oP\,u = 0$ as well as $\oP^*\,\oP\,u = 0$ are covariant equations. Only the latter one has 
a 'common' Lagrange density, namely $\L_\oP= \frac{1}{2}\,(\oP\,u)^2$, which contains only pdo's.  
\end{exam}

\begin{exam}\label{subsec:advdiff}
A process modeling advection with constant velocity $v_0$ and diffusion with constant Diffusivity $D_0$ 
can be modeled by   
\begin{equation*}
  \oA\,u = u_t  + \v\cdot\nabla u - D_0\,\Delta\,u = f \quad\mbox{on}\quad \R^4 
\end{equation*}
with $\v \equiv (v_1,v_2,v_3)$.
A covariant formulation of this advection-diffusion equation is obtained if $\oA$ is written as follows 
$$
    \oA = v_\mu \partial^\mu - D_0 \partial_\mu \partial^\mu + D_0 \partial_0 \partial^0 
    \quad\mbox{with}\quad
    \mbox{$c=1$ in $\partial_0$, $v_0 := -1$.}
$$
From the equivalent equation $\oA^\#\oA\,u=\oA^\# f$ with 
$\oA^\# := v_\mu \partial^\mu + D_0 \partial_\mu \partial^\mu - D_0 \partial_0 \partial^0$, i.e. 
$$
   (v_\mu \partial^\mu)^2\,u + D_0^2(\partial_\mu \partial^\mu - \partial_0 \partial^0)^2\,u = \oA^\#\,f \,,
$$
we infer the following pdo-Lagrange density  
\begin{equation*}
    \L
       = \frac{1}{2}\,(v_\mu \partial^\mu u)^2 
      - \frac{1}{2}\,D_0^2\,(\partial_\mu \partial^\mu u - \partial_0 \partial^0 u)^2
      - (\oA^\# f)\,u\,. 
\end{equation*}
\end{exam}

We conclude this section by deriving a covariant Lagrange density of the wave equation of Nachman, 
Smith and Waag for homogeneous media with one relaxation parameter. 

\begin{exam}
The wave equation of Nachman, Smith and Waag for inhomogeneous media with one relaxation parameter is 
given by $\tilde \oA\,u = \tilde f$, where 
$$
  \tilde\oA\,u :=
    \frac{\tau_0}{\rho\,c_0^2}\,u_{ttt} 
      + \frac{1}{\rho\,c_0^2}\,u_{tt} 
      - \nabla\cdot\left( \frac{\nabla u}{\rho} \right)       
      - \tau_1\,\nabla\cdot\left( \frac{\nabla u_t}{\rho} \right) 
      = 0
$$
and $\tilde f$ models the initial conditions 
$$
    u(\x,0+) = u_0(\x)\,,\quad u_t(\x,0+) = u_1(x) \quad\mbox{and}\quad u_{tt}(\x,0+) = u_2(\x)\,.
$$
Here $c_0$, $\rho$ and $\tau_1$ denote a characteristic sound speed, the equilibrium density 
and the relaxation time, respectively. (Each of these function can depend on $\x$.) 
$\tau_0$ is some function of $\x$ that depends on $c_0$, $\tau_1$, $\rho$ and the 
compressibility. We think that this is the most complicated equation in this paper, 
but it is also the most interesting one.

To simplify the matter, we restrict us to the homogeneous case and set $c_0=1$. Then the density can be canceled out 
and we get $\oA\,u=f$ with 
$$
  \oA := \tau_0\,\partial_t^3  + \partial_t^2 - \left(\Id + \tau_1\,\partial_t\right)\,\Delta  
  \qquad \mbox{($\Id$ identity)}\,.
$$
Here $f$ models the above initial conditions. 
A convariant formulation of this operator is given by 
$\oA_{c} := (\tau_0-\tau_1)\,\partial_0^3 - \left[\Id + \tau_1\,\partial_0\right] \,\partial_\mu \partial^\mu$.  
It is convenient to reorganize its terms as follows 
$$
  \oA_{c} 
     = [ (\Delta\tau)\,\partial_0^3 - \tau_1\,\partial_0\,\partial_\mu \partial^\mu ] 
         - \partial_\mu \partial^\mu  
        \qquad\mbox{with}\qquad 
    \Delta\tau:=\tau_0-\tau_1\,.
$$
The adjoint of this operator, i.e. 
$\oA_{c}^*:= - [ (\Delta\tau)\,\partial_0^3 - \tau_1\,\partial_0\,\partial_\mu \partial^\mu ] - \partial_\mu \partial^\mu$,
is also covariant and its application to the above equation yields the (equivalent) normal equation\footnote{The 
equivalence follows for example from the Fourier transform.} 
$$
  -[ (\Delta\tau)\,\partial_0^3 - \tau_1\,\partial_0 \,\partial_\mu \partial^\mu  ]^2\,u
         + (\partial_\mu \partial^\mu)^2 \,u  
      = \oA_c^*\,f \,,
$$
which is covariant. The respective covariant Lagrange density is given by 
$$
   \L_{NSW} 
        =  \frac{1}{2}\,[ (\Delta\tau)\,\partial_0^3\,u - \tau_1\,\partial_0 \,\partial_\mu \partial^\mu\,u  ]^2
           + \frac{1}{2}\,(\partial_\mu \partial^\mu\,u)^2 
           - (\oA_c^*\,f)\,u \,
$$
that contains only pdo's. The above normal equation is a pde of order $6$ and requires three additional 
initial conditions
\begin{equation*}
\begin{aligned}
   (\partial_t^3 u)(\x,0+) = u_3(\x)\,,\quad  
   (\partial_t^4 u)(\x,0+) = u_4(\x) \quad\mbox{and}  \quad 
   (\partial_t^5 u)(\x,0+) = u_5(\x)\,.
\end{aligned}
\end{equation*}
Because $(\oA\,u)(\x,0+)=0$ for $t>0$, it follows that $\tau_0\,u_3 + u_2 - \Delta\,u_0 - \tau_1\,\Delta\,u_1 = 0$ 
and thus
$$
  u_3 = - \frac{1}{\tau_0}\,u_2 + \frac{1}{\tau_0}\,\Delta\,u_0 + \frac{\tau_1}{\tau_0}\,\Delta\,u_1\,.
$$
Similarly, $(\oA\,u_t)(\x,0+)=0$ and $(\oA\,u_{tt})(\x,0+)=0$ for $t>0$ imply 
$$
  u_4 = - \frac{1}{\tau_0}\,u_3 + \frac{1}{\tau_0}\,\Delta\,u_1 + \frac{\tau_1}{\tau_0}\,\Delta\,u_2
\quad
\mbox{and} 
\quad 
  u_5 = - \frac{1}{\tau_0}\,u_4 + \frac{1}{\tau_0}\,\Delta\,u_2 + \frac{\tau_1}{\tau_0}\,\Delta\,u_3\,.
$$
\end{exam}

\begin{rema}
From the last example, we see that $\L_{NSW} = \L_0 + \L_1$, where 
\begin{itemize}
\item  $\L_0 := \frac{1}{2}\,(-\partial_\mu \partial^\mu\,u)^2 + (\partial_\mu \partial^\mu\,f)\,u$ is a 
          Lagrange density in the absence of dissipation and 
\item  $\L_1 :=  \frac{1}{2}\,(\oA_1\,u)^2 - (\oA_1^*\,f)\,u$ with 
          $\oA_1 :=  (\Delta\tau)\,\partial_0^3 - \tau_1\,\partial_0 \,\partial_\mu \partial^\mu$ 
          incorporates dissipation. 
\end{itemize}
$\L_0$ and $\L_1$ are the Lagrange density of 
$$
  (-\partial_\mu \partial^\mu)^2 \,u = - \partial_\mu \partial^\mu\,f
      \qquad \mbox{and} \qquad 
   \oA_1^*\,\oA_1\,u=\oA_1^*\,f
$$
and $\L_{NSW}$ is the Lagrange density of 
$$
  (-\partial_\mu \partial^\mu)^2 \,u + \oA_1^*\,\oA_1\,u 
      = \oA_1^*\,f - \partial_\mu \partial^\mu\,f \,,
$$
respectively. 
This equation is formally a superposition of the previous ones, which is very convenient for modeling dissipation.  
The same is true for the Telegraph equation if $\oA_1 := d_0\,\partial_0$. 
\end{rema}

\section{Further examples}
\label{sec-FurEx}

\subsection{Shear wave equation}
\label{subsec-ShearWave}

We now derive a (non-trivial) Lagrange density of the \emph{shear wave equation}. If $\u$ denotes the 
\emph{vector displacement}, then it is well-known that there exists a \emph{scalar-valued function} 
$\psi$ (potential) such that $\u := \nabla \psi$ holds. The shear wave equation for $\psi$ reads as follows 
$$
    \psi_{tt} - c_0^2\,\Delta \psi_t = 0   \qquad \mbox{(diffusion like)}\,. 
$$ 
We endow it with the initial data $\psi|_{t=0}=\psi_1$ and $\psi_t|_{t=0}=\psi_2$. This equation is nothing else 
but the time derivative of the diffusion equation. We recall that a Lagrange density of the diffusion equation 
(endowed with initial conditions) reads as follows
$$
  \L_{diff} = \frac{1}{2}\,u_t^2 + \frac{1}{2}\,\left[ \nabla\cdot (D\nabla u) \right]^2  
        \quad\mbox{for}\quad t>0\,. 
$$
With a trick we can employ this Lagrange density. Namely, if $\psi$ solves the shear wave equation, then 
$$
     \psi_t - c_0^2\,\Delta \psi = a  
     \qquad\quad\mbox{for some function $a$ of $\x$.}
$$
\textbf{Case 1.} If $a=0$, then this is just the diffusion equation with diffusivity $D:=c_0^2$. As it should be,  
it is positive. Hence we know its Lagrange density and its Energy-Hamiltonian, where the latter is just the 
zero function. \\
Is it possible that $a=0$? 
To answer that question we posed the problem on the whole time domain with a source term modeling 
initial data and demanding the causality condition: 
$\psi(\cdot,t)=0$ for $t<0$? Because the function $a$ depends only on $x$ and not on 
time, we infer from the causality condition that 
$$
     a(\x) = \psi_t(\x,t) - c_0^2\,\Delta \psi(\x,t) = 0 + c_0^2\,0 = 0  
     \quad\quad\mbox{for each $\x\in\R^3$ and $t<0$.}
$$
Hence the function $a$ must vanish. 
This is a good example for demonstrading that it may be crucial how a problem is posed. 
We actually prefer to pose problems on the whole time line, then the Fourier tranform can be used to show uniqueness 
and much more. (And then we can also use fractional derivatives.)\\ 
\textbf{Case 2.} Now let $a\not=0$. From the equivalent equation $\oA^\#\oA\,u=0$ with 
$\oA^\# := \partial_t + c_0^2\,\Delta$, i.e. 
$$
     \psi_{tt} - c_0^4\,\Delta^2\psi = c_0^2\,\Delta a   
     \qquad\mbox{with}\qquad
     \psi|_{t=0}=\psi_1  \quad\mbox{and}\quad
     \psi_t|_{t=0}=\psi_2\,,
$$
we infer the Lagrange density 
$$ 
  \L 
    = \frac{1}{2}\,\psi_t^2 
      + \frac{1}{2}\,c_0^4\,(\Delta\psi)^2 
      + c_0^2\,\Delta a\,\psi \,.
$$
Because this Lagrange density implies the equation $\psi_{tt} - c_0^2\,\Delta \psi_t = c_0^2\,\Delta a$, 
we infer for the function $a$ the condition $\Delta a = 0$. Then the last term in the previous 
Lagrange density drops out and we arrive at the same Lagrange density as in the first case. This is also a 
demonstration of the usefulness of our approach.

\subsection{The Beam equation}

Let us now investigate the \emph{Beam equation} $\oA\,u = u_t + u_{xxxx} = f$ with $f(x,t) = \varphi(x)\delta(t)$. 
Then $u(x,0+)=\varphi(x)$. We assume that $\varphi$ vanishes for sufficiently large $x$. 
Its equivalent normal equation is given by 
$$
        [\partial_t^2 - \partial_x^8] \,u = [\partial_t - \partial_x^4]\,f
$$
and implies the Lagrange density 
$$
      \L = \frac{1}{2}\,u_t^2 + \frac{1}{2}\,u_{xxxx}^2 + ([\partial_t - \partial_x^4]\,f)\,u \,.
$$
Similarly as for diffusion, the Energy-Hamiltonian $\H_E$ vanishes (if $f$ models initial data.) However, 
the Mass-Hamiltonian is positive and conserved, indeed we have 
$$
      \frac{\d H_M}{\d t}[u] 
        = \int_\R \oA^{-1}\,f_t \, \d x 
        = - \int_\R u_{xxxx} \, \d x 
        = -\lim_{z\to \infty} \varphi_{xxx}|_{x=-z}^{x=z}
        = 0
$$
Moreover, it follows that $\frac{\d H_M}{\d t}[\partial_t^n u] = 0$ for $n\in\N$ and thus  
$\int_{\R} \partial_x^{3 n} u \,\d x = 0 \quad\mbox{for}\quad n\in\N$ (if $u$ is sufficiently smooth 
such that $\partial_x^{3 n} u$ exists).

\subsection{A fractional problem}
\label{subsec-FracExam}

In this section, we apply the principle of least action to a \emph{fractional equation} of \emph{rational order}. 
We consider the problem 
\begin{equation}\label{fraceq01}
      [\partial_t^{ \frac{1}{2} } u + \nabla \cdot(\v \, u) ](\x,t) 
         = \varphi(\x)\,(\partial_t^{ -\frac{1}{2} } \delta)(t) =: f(\x,t)
          \quad \mbox{for}\quad (\x,t)\in\R^4  
\end{equation}
with $u|_{t<0} = 0$ and $\v=(v_1,v_2,v_3)\in\R^3$ (constant). Here $\partial_t^{ \alpha }$ denotes the 
\emph{Riemann Liouville fractional derivative} of order $\alpha>0$ defined by (cf.~\cite{KiSrTr06,Ma10,GaWi99})
$$
   \F\{ \partial_t^{ \alpha } u \}(\omega) = (-\i\,\omega)^\alpha  \, \F\{u\}(\omega)  \qquad 
   \mbox{ ($\F$ Fourier transform) }\,.
$$
We assume that $\varphi$ vanishes for sufficiently large $|\x|$. 
This problem is equivalent to 
\begin{equation}\label{fraceq02}
      [\partial_t u + \nabla \cdot(\v \, \partial_t^{ \frac{1}{2} }u) ](\x,t) 
         = \varphi(\x)\, \delta(t)  
          \quad \mbox{for}\quad (\x,t)\in\R^4  \,,
\end{equation}
which shows that $\varphi$ is nothing else but the initial data, i.e. we have $u(\cdot,0+) = \varphi$. 
This justifies our choice of source term in~(\ref{fraceq01}).

For convenient, we define $\oB := \nabla \cdot(\v \quad)$. In a second step, we isolate the fractional derivative in  equation~(\ref{fraceq01}) and insert the result in equation~(\ref{fraceq02}) and obtain the 
diffusion equation 
$$
    \partial_t u - \oB^2\,u 
        = \varphi\,\delta + f_1 \quad\mbox{,where}\quad 
        f_1 := - (\oB \varphi) \, (\partial_t^{ -\frac{1}{2} } \delta) \,.
$$  
In this example, the \emph{memory part} is entirely described by $f_1$. 
Because $\partial_t^{ -\frac{1}{2} }$ is not a local operator, in general, $f_1$ does not 
vanish for positive time. 
According to our previous results, a Lagrange density is given by 
\begin{equation*}
\begin{aligned}       
   \L(u) 
      = \frac{1}{2} (\partial_t u)^2 + \frac{1}{2} (\oB^2 u)^2 
         + u\,(\partial_t + \oB^2) f_1  
         + g_0 \, u \,,
\end{aligned}
\end{equation*}
where $g_0 := \varphi\,\delta' + (\oB^2 \varphi)\,\delta$ models the initial data $u(\cdot,0+) = \varphi$ and 
$u_t(\cdot,0+) = \oB^2 \varphi$. This Lagrange density depends explicit on time. 
For $t>0$ the Mass-Hamiltonian satisfies  
\begin{equation*}
\begin{aligned}
 \frac{\d H_M}{\d t} (t)
        = \int_\R f_1(\x,t) \, \d \x  
        = - (\partial_t^{ -\frac{1}{2} } \delta)(t) \int_\R \nabla\cdot (\v \, \varphi)(\x,t) \,  \, \d \x 
        = 0 
\end{aligned}
\end{equation*}
and thus if we interpret $u$ as a density, the mass is conserved.  
For positive time the usual Hamiltonian density reads as follows
\begin{equation*}
\begin{aligned}
 \H 
     = \frac{\partial \L}{\partial u_t}\,u_t - \L 
     = \frac{1}{2} (\partial_t u)^2 
         - \frac{1}{2} (\oB^2 u)^2 
         - u\,(\partial_t + \oB^2) f_1\,,
\end{aligned}
\end{equation*}
which is not conserved and does not vanish. 
In summary, we have shown that this problem is equivalent to a diffusion problem with special source term 
that models memory. This process conserves the mass but not $H$. 

We note that if the above strategy is applied to 
$\partial_t^{ \frac{1}{n} } u + \nabla \cdot(\v \, u)  =  \varphi\,(\partial_t^{ -\frac{1}{n} } \delta)$, 
where the second step is repeatedly applied, a similar result is obtained. For general rational orders the situation 
is more complicated and lies beyond the scope of this paper.

\section{Conclusions}
\label{sec-Concl}

In this paper, it has been shown that under fairly weak conditions for linear pde's and linear fractional equations 
(covariant) Lagrange densities exist (e.g. the trivial one). Moreover, if the pde has constant coefficients, 
then a covariant pdo-Lagrange density exists. In particular, we have derived and discussed covariant Lagrange 
densities for several well-known dissipative equation for which a Lagrange density was not known (or was believed 
not to exist). In our approach, each  Lagrange density (for a given set of variables) implies an Hamiltonian, 
which is conserved if the considered problem is of a certain type. We have shown that the usual Lagrange density 
implies an Hamiltonian that models the energy or is part of an Hamiltonian that models the energy. 
It is fairly clear that there are many unknown conserved Hamiltonians (and higher order Hamiltonians) 
out there. Moreover, as demonstrated by two examples, the principle of least action is also applicable and very useful 
for time-fractional problems. A new field of research seems to be opened which is very useful for physics 
as well as mathematics. It leads to a more unified point of view and permits an extended handling of the 
principle of least action. We think that the expertise from several different areas in physics will shine 
a much brighter light on it.

\end{document}